\documentclass[aps,prl,showpacs,twocolumn,superscriptaddress]{revtex4-2}%

\usepackage{graphicx}
\usepackage{dcolumn}
\usepackage{bm}

\usepackage{float}  

\usepackage{graphicx}
\usepackage{graphics}
\usepackage[caption=false]{subfig}
\usepackage{color}
\usepackage[english]{babel}
\usepackage[utf8]{inputenc}
\usepackage{ae}
\usepackage{latexsym}
\usepackage{dcolumn}
\usepackage{bm}
\usepackage[table]{xcolor}
\usepackage{array}
\usepackage{url}
\usepackage{epsf}
\usepackage{epsfig}
\usepackage{pstricks,pst-grad}
\usepackage[pdftex,colorlinks]{hyperref}
\usepackage{amsmath}
\usepackage{amssymb}
\usepackage{ragged2e}
\usepackage{comment}

\begin{document}

\title{Electrostatic Correlation Augmented Self-Consistent Field Theory and Its Application to Polyelectrolyte Brushes}

\author{Chao Duan}
\affiliation{Department of Chemical and Biomolecular Engineering, University of California Berkeley, CA 94720, USA}

\author{Nikhil R. Agrawal}
\affiliation{Department of Chemical and Biomolecular Engineering, University of California Berkeley, CA 94720, USA}

\author{Rui Wang}
\email {ruiwang325@berkeley.edu}\affiliation{Department of Chemical and Biomolecular Engineering, University of California Berkeley, CA 94720, USA}
\affiliation{Materials Sciences Division, Lawrence Berkeley National Lab, Berkeley, CA 94720, USA}

\date{\today}

\begin{abstract}
Modeling ion correlations in inhomogeneous polymers and soft matters with spatially varying ionic strength or dielectric permittivity remains a great challenge. Here, we develop a new theory which systematically incorporates electrostatic fluctuations into the self-consistent field theory for polymers. Applied to polyelectrolyte brushes, the theory predicts that ion correlations induce non-monotonic change of the brush height: collapse followed by reexpansion. The scaling analysis elucidates the competition between the repulsive osmotic pressure due to translational entropy and the attraction induced by ion correlations. We also clarify the absence of causal relationship between the brush collapse-reexpansion and the inversion of the surface electrostatic potential. Furthermore, strong ion correlations can trigger microphase separation, either in the lateral direction as pinned micelles or in the normal direction as oscillatory layers. Our theoretical predictions are in good agreement with the experimental results reported in the literature.

\end{abstract}

\maketitle


Ion-containing polymers and soft matters with spatially varying ionic strength or dielectric permittivity are ubiquitous \cite{Decher_1997,Stradner_2004,Kim_2020,Galvanetto_2023}. One example is polyelectrolyte (PE) brushes grafted on substrates which regulate surface properties such as transport, wettability, adhesion, antifouling and lubrication \cite{Xiong_2023,Xin_2010,Stuart_2010,Shabani_2024,Raviv_2003,Chen_2009}.  PE brushes are very sensitive to the addition of salt ions \cite{R_he,Romet-Lemonne:2004wx}. The primary role of monovalent ions, screening and translational entropy, is well captured by the mean-field Poisson-Boltzmann theory \cite{Messina_2009,dean2014electrostatics}.
However, brush morphology and mechanical behaviors in  multivalent salt solutions are quite non-trivial \cite{Mei:2006vx,Kegler:2008ue,Fazli_2006}.
Using neutron reflectivity and surface forces apparatus, Tirrell group found that the onset of brush collapse is reduced by 3-6 orders of magnitudes in multivalent ions compared to monovalent ions \cite{Yu_2016}. It was observed that brush height changes non-monotonically: collapse followed by reexpansion as salt concentration further increases. Similar non-monotonic size change was reported in the study of single PE conformation \cite{Hsiao:2006te}. AFM images showed that trivalent ions induce lateral structural inhomogeneities of PE brushes, which were further confirmed by molecular simulations \cite{Yu_2017,Jackson_2017}. Furthermore, Yu et al. also reported that the friction force between two opposing PE brushes increases dramatically in multivalent salt solutions, diminishing the lubricity \cite{Yu_2018}. These phenomena cannot be even qualitatively explained by mean-field electrostatics. 

An obvious factor missing in the mean-field theories is the electrostatic correlation \cite{Luo_2006,Bakhshandeh_2011,Gupta_2020}. Despite many efforts, it remains a great challenge to model the coupling between electrostatic correlation and chain conformation particularly in inhomogeneous systems.
Muthukumar developed a variational method based on a two-state model for counterion binding \cite{Muthukumar_2004,Kundagrami_2008}.
This theory pre-integrates the salt degrees of freedom, resulting in a screened interactions between charged monomers, which ignores the feedback on the correlations between ions.
Applying this method to PE brushes, Li et al. explained the sharp collapse in multivalent solutions but failed to capture the subsequent reexpansion \cite{Li_2022,Li_2024}. 
Shen and Wang addressed the electrostatic fluctuations coupled with chain connectivity through quantifying the self-energy of a PE \cite{Shen_2017,Shen_2018}.
Recently, Sing and Qin treated the correlation via cluster expansions \cite{Sing_2023}.
However, all three approaches invoked preassumed functional form of chain conformation, limiting their application to PEs with various structures like globule and pearl-necklace \cite{DOBRYNIN20051049,Duan_2020}. These theories are also not applicable to inhomogeneous systems.
Furthermore, a hybrid liquid state integral equation and self-consistent field theory (SCFT) was developed to model correlations in inhomogeneous polymer blends and block copolymers \cite{Sing:2013aa,Sing_2014}. Local charge neutrality is assumed, which ignores long-range components of electrostatics. This assumption fails wherever a double layer with local charge-separation exist, e.g. PE adsorption and brushes \cite{Gurovitch_1999,Dobrynin_2000,Winkler_2006,Li_2006,Mei:2006vx,Kegler:2008ue,Yuan_2024_ChargeRegulation,Yu_2016,Yu_2017,Yu_2018,Jackson_2017,Hsiao:2006te}.
Fredrickson group embedded fluctuation into SCFT through one-loop expansion of the Hamiltonian to second order around the saddle point. However, the non-perturbative feedback of electrostatic fluctuations has not been included, which is essential to explain the non-trivial salt effects \cite{Audus_2015}.

We develop a new theoretical framework which systematically incorporates the variational Gaussian renormalized fluctuation theory for electrostatics \cite{Wang:2010wk,Agrawal:2022ux,Agrawal_2022,Agrawal_2023,Agrawal_2024} into SCFT for polymers \cite{Duan_2020,Fredrickson_Book2006,Duan:2023vk}. The coupling between ion correlation and chain conformation can be fully captured without any preassumption on the PE structure itself. Our theory is particularly capable of modeling inhomogeneous PEs with spatially varying ionic strength or dielectric permittivity and can be performed in high-dimensional calculation. Applied to PE brushes, the theory predicts the non-monotonic salt concentration dependence of brush height and lateral microphase separation, in good agreement with experiments reported in the literature.


We consider a general system consisting of $n_P$ negatively charged PEs each with $N_P$ monomers in an electrolyte solution which contains $n_S$ solvent molecules, $n_C$ monovalent counterions ($z_C=+1$) and $n_{\pm}$ salt ions with valency $z_{\pm}$. The system is taken as a semicanonical ensemble: the number of PEs and counterions are fixed whereas solvents and salt ions are connected with a bulk salt solution of ion concentration $\rho^b_{\pm}$ that maintains the chemical potentials of $\mu_S$ and $\mu_{\pm}$. To more accurately describe the charge interactions, the ionic charge is modeled by a finite spread function $h_K({\bf r},{\bf r}^{\prime})$ ($K= P, C,\pm$) instead of using the point-charge model \cite{Wang:2010wk,Agrawal_2022,Agrawal:2022ux,Agrawal_2023,Agrawal_2024}. For convenience, we take $h_K({\bf r},{\bf r}^{\prime})$ to be Gaussian which retains the Born radius $a_{K}$. The smeared charge model for PE widely used in existing theories fails to capture any correlations between charged monomers as well as between them and mobile ions. We adopt a discrete Gaussian chain model where each monomer is a charged particle with valency $z_P$. The number of dissociated counterions is thus $n_C=-z_P n_P N_P$.
The partition function is
\begin{equation}\label{PartitionFunc}
\begin{aligned}
\mathcal{Z} &= \frac{1}{n_P!v_P^{n_P N_P}} 
\prod_{i=1}^{n_P} \int \hat{D}\{ {\bf R}_i \} \frac{1}{n_C!v_C^{n_C}}  \prod_{j=1}^{n_C} \int d{\bf r}_{C,j} \\
& \cdot \sum_{\gamma=S,\pm} \sum_{n_\gamma = 0}^{\infty} \frac{e^{\mu_\gamma n_\gamma}}{n_\gamma! v_\gamma^{n_\gamma}} \prod_{k=1}^{n_\gamma} \int d{\bf r}_{\gamma,k} {\rm exp}\left(-H \right)
\end{aligned}
\end{equation}
where $v_P$ and $v_K$ ($K= S, C,\pm$) are the volume of the monomers and small molecules, respectively.
We assume $v_P=v_S=v$. $\int \hat{D}\{ {\bf R}_i\}$ denotes integration over all configurations weighted by the Gaussian-chain statistics \cite{Fredrickson_Book2006}. The integrals in Eq. \ref{PartitionFunc} are under the incompressibility constraint $\sum_{K} \hat{\rho}_K v_K =1$ for all species where $\hat{\rho}_K$ is the instantaneous number density of species $K$. The Hamiltonian
\begin{equation}\label{Hamiltonian}
\begin{aligned}
H=\chi v \int d{\bf r}
\hat{\rho}_P({\bf r}) \hat{\rho}_S({\bf r})
+\frac{1}{2} \int d{\bf r} d{\bf r}^{\prime}
\hat{\rho}_e({\bf r}) C({\bf r},{\bf r}^{\prime}) \hat{\rho}_e({\bf r}^{\prime})
\end{aligned}
\end{equation}
includes hydrophobic interaction between polymer and solvent in terms of Flory-Huggins $\chi$ parameter and the long-range electrostatic interaction between all charged species. $\hat{\rho}_e({\bf r})=\sum_K \int d{\bf r}^{\prime} z_K h_K({\bf r}-{\bf r}^{\prime}) \hat{\rho}_K({\bf r})$ with $K=P, C, \pm$ is the net charge density, and $C({\bf r},{\bf r}^{\prime})$ is the Coulomb operator.

Following the standard field theoretical approach \cite{Fredrickson_Book2006}, the interacting system can be decoupled into noninteracting chains and ions in fluctuating fields (see the detailed derivation in Sec. I of the Supplemental Materials). For the electric field, instead of saddle-point approximation, we use a non-perturbative variational approach to account for fluctuation by introducing a general Gaussian reference action \cite{Wang:2010wk,Agrawal_2022,Agrawal:2022ux,Agrawal_2023,Agrawal_2024,Shen_2017,Shen_2018}.
The consequence of this fluctuation is the
self-energy of charged species. 
For small mobile counter ions and salt ions, the self-energy $u^0_K({\bf r})$ can be individually evaluated from their own charge spread \cite{Agrawal_2022,Agrawal:2022ux,Agrawal_2023,Agrawal_2024}. In contrast, the charge of a polymer spreads the pervaded volume of the entire chain \cite{Shen_2017,Shen_2018}. The polymer self-energy, when evaluated from its constituting monomer $i$, includes both the individual contribution $u^0_P({\bf r})$ and an extra $u^{ex}_P({\bf r};i)$ attributed by all other intra-chain monomers.

The key results of our theory are the self-consistent equations that capture the coupling between charge correlation (manifested by the self-energy $u^0_K ({\bf r})$ and $u^{ex}_P({\bf r};i)$) and chain conformation (described by the propagator $q({\bf r};i)$ and its complementary $q^{\dagger}({\bf r};i)$):
\begin{subequations}\label{u}
\begin{align}\label{u0}
u^0_K({\bf r}) &= \frac{z^2_K}{2} \int d{\bf r^{\prime}} d{\bf r^{\prime\prime}} h_K({\bf r},{\bf r}^{\prime}) G({\bf r}^{\prime},{\bf r}^{\prime\prime}) h_K({\bf r^{\prime\prime}},{\bf r})
\end{align}
\begin{align}\label{u_ex}
&u^{ex}_P({\bf r};i) = \frac{z^2_P}{2}   \biggl[ \sum^{i-1}_{j=1} \frac{\int d{\bf r^{\prime}} q_0({\bf r}^{\prime};j) e^{V^0_P({\bf r}^{\prime})} G({\bf r}^{\prime},{\bf r}) g^{\dagger}_0({\bf r}^{\prime},{\bf r};j,i)}{q_0({\bf r};i)} \nonumber\\
&+ \sum^{N_P}_{j=i+1} \frac{\int d{\bf r^{\prime}} g_0 ({\bf r},{\bf r}^{\prime};i,j) G({\bf r},{\bf r}^{\prime}) e^{V^0_P({\bf r}^{\prime})} q_0^{\dagger}({\bf r}^{\prime};j)}{q_0^{\dagger}({\bf r};i)} \biggl]
\end{align}
\end{subequations}
\begin{subequations}\label{q}
\begin{align}\label{q_forward}
q({\bf r};i) &= e^{-V_P({\bf r};i)} \int d{\bf r^{\prime}} \Phi ({\bf r} - {\bf r^{\prime}}) q({\bf r^{\prime}};i-1)
\end{align}
\begin{align}\label{q_backward}
q^{\dagger}({\bf r};i) &= e^{-V_P({\bf r};i)} \int d{\bf r^{\prime}} \Phi ({\bf r} - {\bf r^{\prime}}) q^{\dagger}({\bf r^{\prime}};i+1)
\end{align}
\end{subequations}
$G({\bf r},{\bf r}^{\prime})$ in Eq. \ref{u} is the charge correlation function:
\begin{equation}\label{Green}
\begin{aligned}
-\nabla_{\bf r} \cdot[\epsilon&({\bf r}) \nabla_{\bf r} G({\bf r},{\bf r}^{\prime})] + 2I_0({\bf r}) G({\bf r},{\bf r}^{\prime}) \\
& + \int d{\bf r^{\prime\prime}} 2I_{ex}({\bf r},{\bf r^{\prime\prime}}) G({\bf r^{\prime\prime}},{\bf r}^{\prime})   = \delta({\bf r}-{\bf r}^{\prime})
\end{aligned}
\end{equation}
where $\epsilon({\bf r})$ is the scaled permittivity depending on the local composition of the system \cite{Sing_2014,Wang_2008,Zhuang_2021,Duan_2024}.
$I_0({\bf r})=\sum_{K} z^2_K \rho_K({\bf r})/2$ is the local component of ionic strength. Different from the commonly used Green function, the ionic strength of PE systems has an extra non-local contribution $I_{ex}({\bf r},{\bf r^{\prime}})$ from the long-range charge spread of polymer (see Sec. I in the Supplemental Material). $\Phi ({\bf r} - {\bf r^{\prime}})$ in Eq. \ref{q} is the Gaussian bond transition probability \cite{Fredrickson_Book2006}.
$V_P({\bf r};i)=V^0_P({\bf r})+u^{ex}_P({\bf r};i)$ is the total interaction field experienced by monomer $i$, where $V^0_P({\bf r})=\omega_P+z_P\psi+u^0_P$. $\omega_P= \chi \phi_S + \xi - (\partial \epsilon/\partial \rho_P)(\nabla\psi)^2/2$ is the field conjugate to monomer density with $\xi$ reinforcing the incompressibility.
When evaluating $u^{ex}_P({\bf r};i)$ in Eq. \ref{u_ex}, we expand the Hamiltonian based on a reference state where charged monomers are treated isolatedly that neglects charge correlations from other intra-chain monomers. The chain propagator $q_0({\bf r};j)$ and intra-chain correlation function $g_0({\bf r},{\bf r}^{\prime};i,j)$ are thus subject to the same recurrence relation (Eq. \ref{q}) as $q$ but with interaction field $V^0_P$ corresponding to the reference state.

The electrostatic potential $\psi$ satisfies the Poisson equation $-\nabla\cdot[\epsilon({\bf r})\nabla\psi({\bf r})] = \sum_K z_K \rho_K({\bf r})$. The monomer density $\rho_P({\bf r})=(n_P/Q_P) \sum_{i=1}^N q({\bf r};i) e^{V_P({\bf r};i)} q^{\dagger}({\bf r};i)$ with $Q_P=\int d{\bf r} q({\bf r};N_P)$.
Counterion density $\rho_C({\bf r}) = n_C {\rm exp}(-z_C \psi-u^0_C-v_C \xi/v)/ Q_C$ with $Q_C= \int d{\bf r} {\rm exp}(-z_C \psi-u^0_C-v_C \xi/v)$. Salt ion densities $\rho_{\pm}({\bf r}) =\lambda_{\pm} {\rm exp}(-z_{\pm}\psi-u^0_{\pm}-v_{\pm}\xi/v)$ with the fugacity $\lambda_{\pm}=e^{\mu_{\pm}}/v_{\pm}$ determined by the bulk concentration $\rho^b_{\pm}$.

For a given chain conformation, the correlation between charged species can be calculated from Eq. \ref{Green}, leading to an update of the self-energy (Eq. \ref{u}). This will in turn feedback to the polymer density and chain conformation through Eq. \ref{q}. The coupling between ion correlations and chain conformation is hence self-consistently achieved via this iteration. The numerical details are provided in Sec. II of the Supplemental Material \cite{FFTW,Hoffman_2018}.



The theory is general for all homogeneous and inhomogeneous PE systems. Here we apply it to PE brushes where one end of each chain is tethered to a planar surface. Without loss of generality, we consider negatively charged brushes immersed in a $z_+$:1 salt solution with the bulk salt concentration $\rho_b$. To focus on the ion correlation effects, we assume $\Theta$ solvent ($\chi$=0.5) and neglect the excluded volume of ions ($v_C=v_{\pm}=0$). The dielectric constant is uniformly set to be 80. The contribution of intra-chain correlation to self-energy $u^{ex}_P$ is found less important than the individual contribution of a charged monomer $u^0_P$ under the condition of high ionic strengths (either high salt concentration or high PE density) \cite{Shen_2017}. For systems we are interested in, $u^{ex}_P$ is also neglected.

\begin{figure}[t]
\centering
\includegraphics[width=0.48\textwidth]{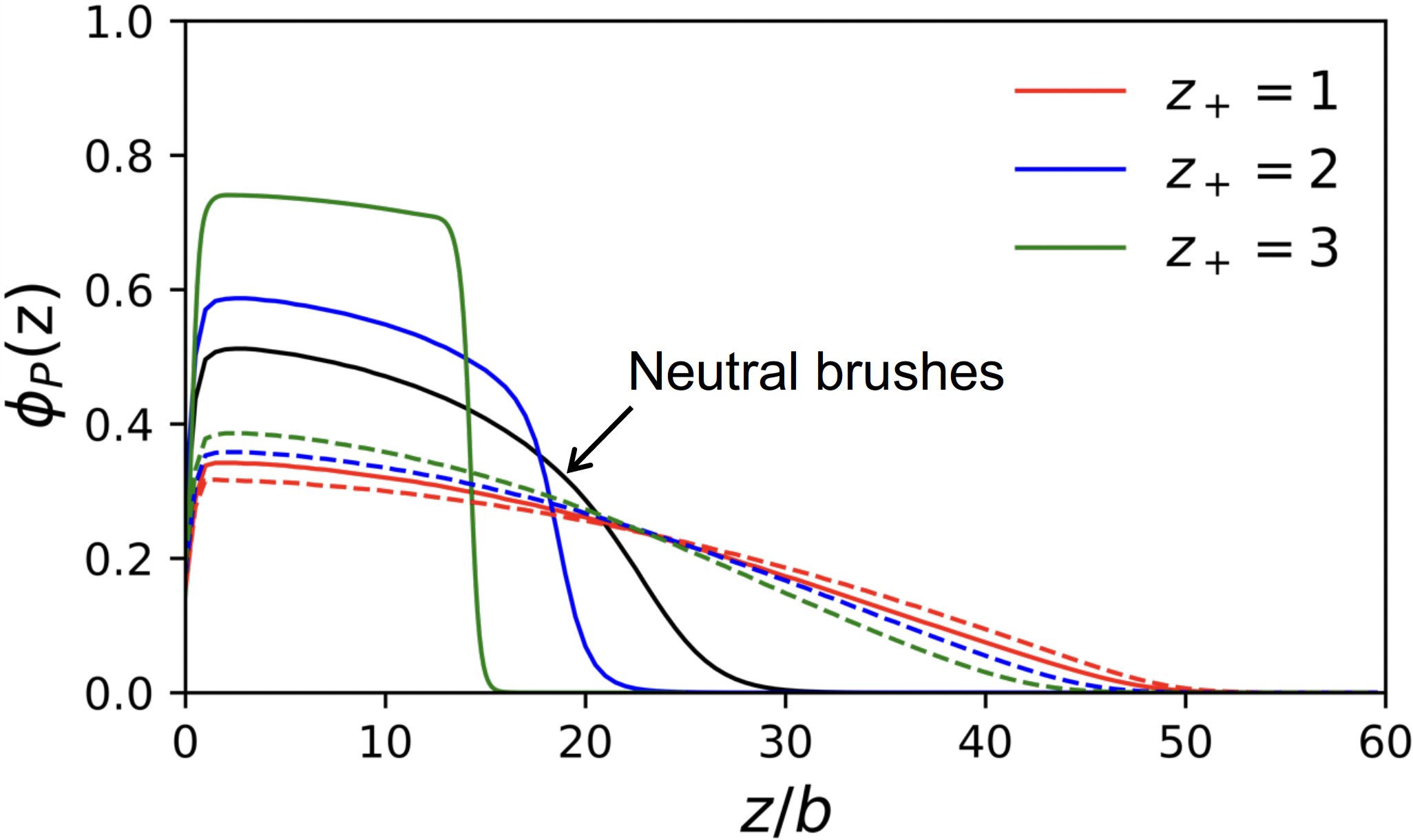}
\caption{The effect of ion correlations on the density profile of PE brushes in the normal direction in comparison with the mean-field results (dashed lines). $\phi_P(z)$ is the polymer volume fraction. $N=100$, $b=1.0$nm, $v=1.0$nm$^3$, $z_P=z_-=-1$, the Born radius $a_P=a_C=a_{\pm}=2.5\mathring{\rm A}$, and the grafting density $\sigma=0.1$nm$^{-2}$. Bulk ionic strength $I_b=0.3$M for all cases.}
\label{fig1}
\end{figure}

\begin{figure}[h]
\centering
\includegraphics[width=0.48\textwidth]{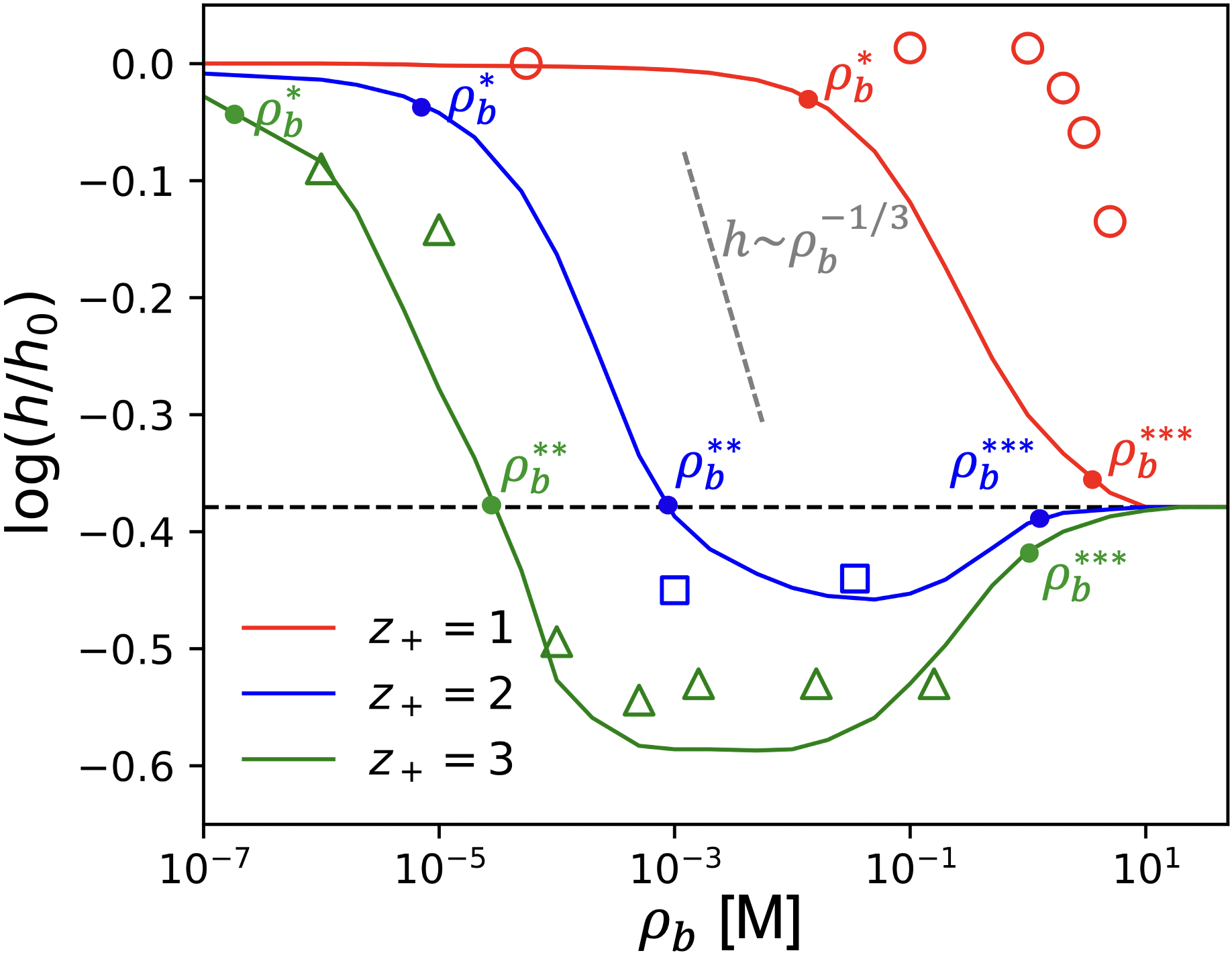}
\caption{Non-monotonic change of brush height on salt concentration. $h$ is normalized by its the salt-free value $h_0$. Lines represent our theoretical predictions and scattering symbols represent experimental
data adopted from Ref. \cite{Yu_2016} with Na$^{+}$ (\textcolor{red}{$\bigcirc$}), Ba$^{2+}$ (\textcolor{blue}{$\square$}) and Y$^{3+}$ (\textcolor{green}{$\triangle$}). The horizontal dashed line is for the charge neutral brushes.
The gray dashed line indicates the scaling relation in the salted regime, $h \sim \rho^{-1/3}_b$.
Critical salt concentrations $\rho^{*}_b$, $\rho^{**}_b$, and $\rho^{***}_b$ are identified on the curves.
$\sigma$ in the calculations has the same value $0.1$nm$^{-2}$ as the experiments.  
}
\label{fig2}
\end{figure}

Electrostatic correlation has dramatic impacts on the brush morphology. Figure \ref{fig1} plots polymer density profiles in the presence of cations with different valency at the same bulk ionic strength. 
The results based on mean-field electrostatics are also provided for comparison, which predict swollen brushes in all cases and insensitivity of the profile on $z_+$. In the stark contrast, with the inclusion of ion correlation, our theory predict that brushes are highly collapsed in multivalent salt solutions. The brush height is even smaller than the charge-neutral case and becomes more compact as $z_+$ increases. This collapse originates from the net attractive force between charged monomers induced by electrostatic correlation.

Our theory fully captures the non-monotonic height change of PE brush in response to multivalent salt. Figure \ref{fig2} plots the brush height $h$ as a function of $\rho_b$, where $h$ is calculated from the Gibbs dividing surface. For $z_+=1$, $h$ shows three regimes consistent with the mean-field results: an ``osmotic regime" at low $\rho_b$ where brushes are highly swollen with a constant $h$, a ``salted regime" at intermediate $\rho_b$ where the swollen brushes shrink due to salt screening, and a ``neutral regime" at high $\rho_b$ where the charges on brushes are almost screened out and $h$ returns to the neutral level \cite{Zhulina_1995}. However, $h$ for $z_+=2$ and $3$ is qualitatively different from the mean-field result. A new ``collapsed" regime appears after the salted regime, where $h$ drops below the neutral brush level, reaches the minimum and reexpands to reenter the neutral regime. Our theoretical predictions are in good agreement with the experimental data reported by Tirrell group \cite{Yu_2016}.

We perform the scaling analysis to elucidate the insights of the non-monotonic height change (see Sec. III in the Supplemental Materials). It is built upon the force balance in the brush region: $f_{\rm ela} +f_{\rm int}+ f_{\rm ion} = 0$ \cite{Zhulina_1995}. $f_{\rm ela} \approx - \sigma h/Nb^2$ is the elastic stress due to chain deformation. $f_{\rm int} \approx w_2 \rho^2_P+w_3 \rho^3_P$ is the non-electrostatic interaction between monomers. $\rho_P=\sigma N/h$; $w_2$ and $w_3$ are the second and third virial coefficients, respectively. $f_{\rm ion}$ is the electrostatic osmotic pressure including ion translational entropy and ion correlation. In the dilute limit, $f_{\rm ion}=\sum_{K=\pm}(\rho_K-\rho^b_K)-\pi/24(\kappa^3-\kappa_b^3)$ with $\kappa$ the inverse of the Debye length \cite{Wang:2010wk}. We identify the critical salt concentrations which locate the boundaries between adjacent regimes (see Fig. \ref{fig2}). The onset of the brush shrink $\rho^{*}_b$ occurs when the ion correlation in the brush becomes comparable to the translational entropy  (i.e. $\kappa^3 \approx \rho_+$), leading to the scaling of $\rho^{*}_b$ on the cation valency as 
\begin{equation}\label{rho_1star}
\begin{aligned}
\rho^{*}_b \sim
    \begin{cases}
     1/\left[z_+(z_++1)\right]^3 & \text{(with ion correlations)}\\
     1/\left[z_+(z_++1)\right] & \text{(mean-field)}
    \end{cases}
\end{aligned}
\end{equation}
Compared with the mean-field result, it clearly indicates a much stronger $z_+$ dependence of $\rho^{*}_b$ in the presence of ion correlations. This scaling explains the experimental observation that the onset of brush shrink occurs at the salt concentration a few orders of magnitudes lower for multivalent ions compared to monovalent ions \cite{Yu_2016}. 

Ion correlations get stronger as $\rho_b$ increases. At the critical concentration $\rho^{**}_b$, the correlation-induced attraction exactly compensates the repulsive pressure from the translational entropy of ions, which leads to $f_{\rm ion}=0$ and hence neutral brush apparently.
In the salted regime and thereafter, we work out that $f_{\rm ion}\approx (\rho^{2}_P/z_{+} \rho_b) \Gamma$, where $\Gamma=1-z_+(\rho_+-\rho_b)/\rho_P$ can be understood as the effective relative charge on the brush after counterion adsorption.
$\Gamma$ satisfies
\begin{equation}\label{Gamma}
\begin{aligned}
\frac{(z_++1)\Gamma-1}{\Gamma-1} &=-\frac{z_+(z_++1)\kappa_b}{16\pi\epsilon} \frac{d u}{d (a\kappa_b)}
\end{aligned}
\end{equation}
$u/(8\pi\epsilon a)$ is the self-energy of monovalent ion with Born radius $a$ in the bulk salt solution \cite{Wang:2010wk}. $\rho^{**}_b$ is thus determined at $\Gamma=0$ through Eq. \ref{Gamma}. It should be noted that Eq. \ref{Gamma} is universal for all charge spread models. If $h$ takes the Gaussian form particularly, $u = 1-(a\kappa_b){\rm exp}(a^2\kappa^2_b/\pi) {\rm erfc}(a\kappa_b /\sqrt{\pi})$ \cite{Agrawal_2022,Agrawal:2022ux,Agrawal_2023,Agrawal_2024}. The right hand side of Eq. \ref{Gamma} reflects the correction from ion correlation, which vanishes at the mean-field level leading to $\Gamma=1/(z_++1)>0$. It shows that overcharging cannot happen in the mean-field electrostatics.    

For $\rho_b>\rho^{**}_b$, the effective charge on the brushes change sign, i.e. $\Gamma<0$. Counterions over compensate the bare backbone charges on polymers.  $f_{\rm ion}$ turns from repulsive to attractive in this regime, which leads to the transition from swollen to collapsed brushes. However, overcharging will not continue increasing; instead, it gets suppressed at high salt concentrations because the strength of ion correlations in bulk also increase, which reduces the driving-force for counterions to migrate to brush region from bulk. At the critical concentration $\rho^{\rm{valley}}_b$ where $\partial \Gamma/\partial \rho_b =0$, overcharging as well as the impact of ion correlation gets maximized. PE brushes are compacted to the lowest height, which is followed by a reexpansion as $\rho_b$ further increases. At the high salt concentration $\rho^{***}_b$, the screening length reduces to a monomer size, i.e. $\kappa_b b \approx 1$. The charges on brushes can be negligible, and the brushes behave neutral again. Using the scaling theory, we obtain the brush height in the four regimes as
\begin{equation}\label{Scaling}
\begin{aligned}
\frac{h}{Nb} \approx
    \begin{cases}
      1 & (\rho_b < \rho^{*}_b) \\
      {\left( \sigma \Gamma/z_+b \right)}^{1/3} \rho_b^{-1/3} & (\rho^{*}_b < \rho_b < \rho^{**}_b)\\
      - (\sigma w_3 z_+/\Gamma b) \rho_b & (\rho^{**}_b < \rho_b < \rho^{***}_b)\\
    (\sigma w_2/b)^{1/3} & (\rho_b > \rho^{***}_b)\\
    \end{cases}
\end{aligned}
\end{equation}

\begin{figure}[t]
\centering
\includegraphics[width=0.48\textwidth]{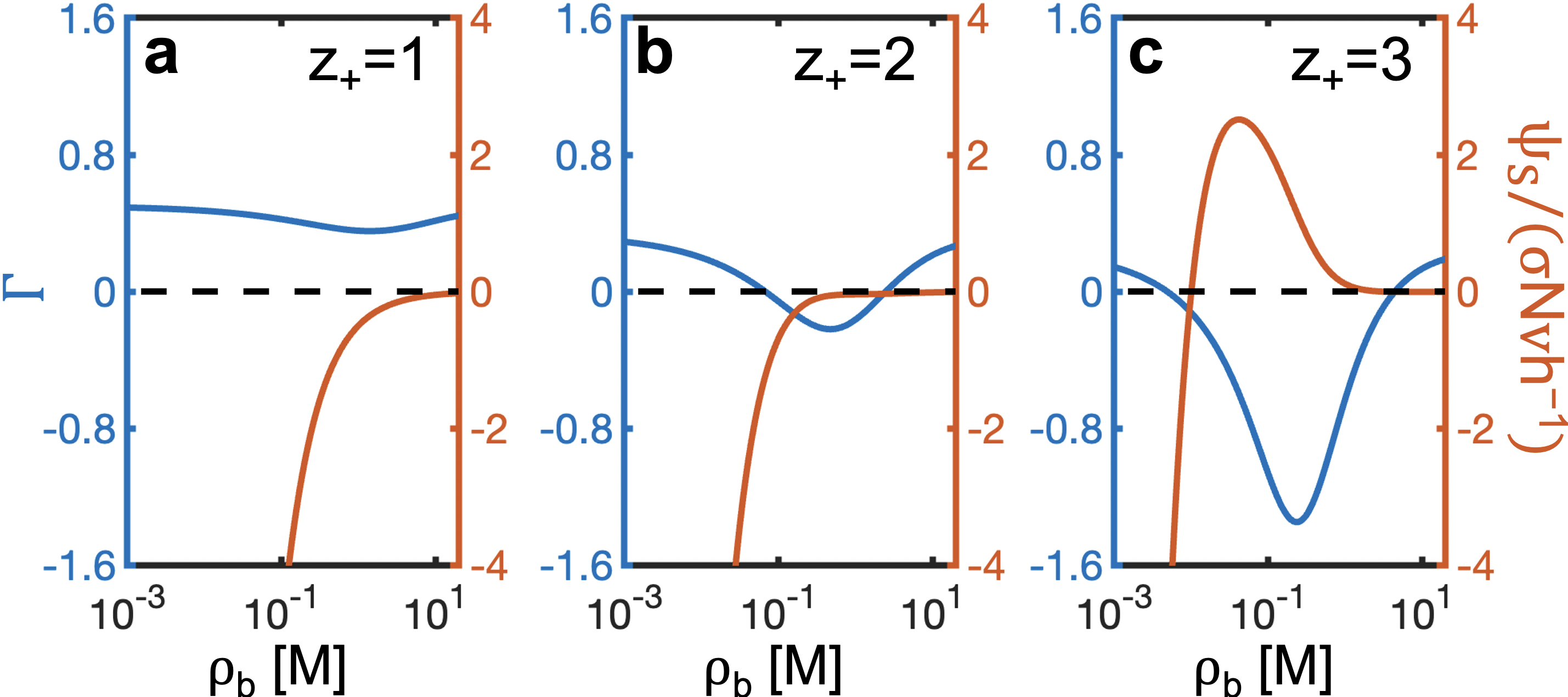}
\caption{The effective relative charge on the brushes $\Gamma$ and the surface electrostatic potential $\psi_{S}$ (normalized by $\sigma N v/h$) vs. $\rho_b$ for cations with different valency.
}
\label{fig3}
\end{figure}

The relationship between chain reexpansion and charge inversion remains under debate \cite{Hsiao:2006te}. The latter in practice is often manifested as the reversal of the surface electrostatic potential $\psi_S$ in electroosmotic flow or electrophoretic mobility \cite{Besteman_2004,Heyden_2006,Agrawal_2024}. In the scaling theory, $\psi_S$ can be approximated by the Donnan potential in the brushes:  
\begin{equation}\label{psi_S}
\begin{aligned}
\psi_S &\approx - \frac{\sigma N}{z^2_+ (z_++1) h \rho_b} \left[ (z^2_+-1)\Gamma+1 \right]
\end{aligned}
\end{equation}
It indicates the absence of charge inversion for $z_+=1$. For multivalent cation, the charge inversion occurs at $\Gamma=-1/(z^2_+-1)$, which is different from the criteria of either the collapse point $\rho^{**}_b$ or the reexpansion point $\rho^{\rm{valley}}_b$. This is clearly illustrated by the salt concentration dependence of $\Gamma$ and $\psi_S$ in Fig. \ref{fig3}. For $z_+=2$, $\Gamma$ changes sign whereas $\psi_S$ remains negative: the brush collapse-reexpansion occurs without charge inversion. For $z_+=3$, both $\Gamma$ and $\psi_S$ change sign; however, they are out of sync and happen at different $\rho_b$. There are hence no causal relationship between the brush collapse-reexpansion and charge inversion, although they are both induced by ion correlation. Our theoretical finding is consistent with the simulation by Hsiao and Luijten on the conformation and mobility of a single PE \cite{Hsiao:2006te}.

\begin{figure}[t]
\centering
\includegraphics[width=0.48\textwidth]{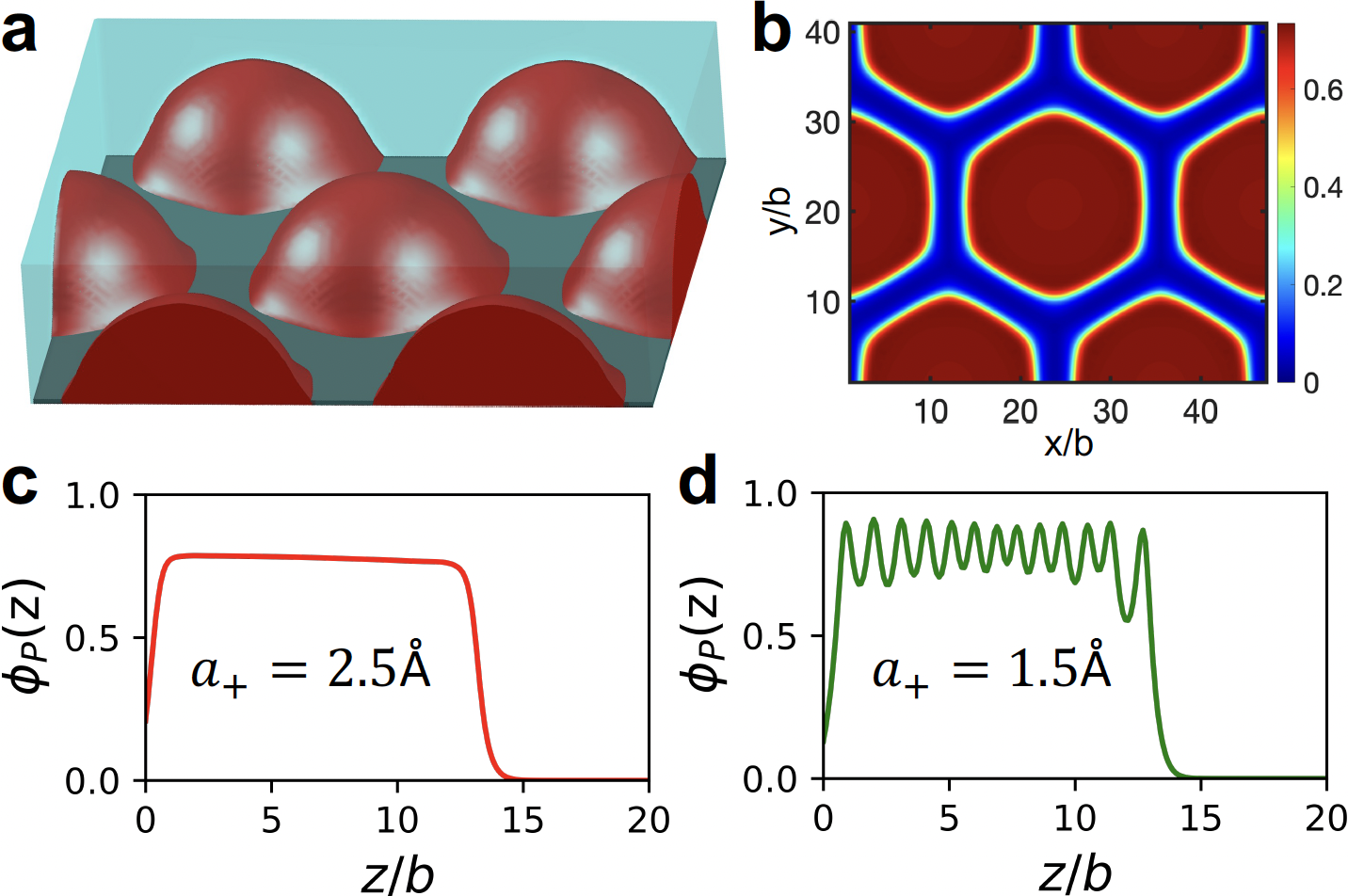}
\caption{Microphase separation in the morphology of PE brushes triggered by trivalent ions. Lateral inhomogeneity of pinned micelles at moderate grafting density $\sigma=0.03 $nm$^{-2}$ presented by (a) 3D isosurface plot and (b) 2D visualization of the polymer density in the $xy$ plane at $z=0$. (c) and (d) plot oscillatory layers formed in the normal direction at high grafting density $\sigma=0.1 $nm$^{-2}$ as cation radius $a_{+}$ decreases from $2.5\mathring{\rm A}$  to $1.5\mathring{\rm A}$. $\rho_b=1$mM.}
\label{fig4}
\end{figure}

Ion correlations can have more profound impacts on the brush morphology besides the apparent height change. At a moderate grafting density $\sigma=0.03$nm$^{-2}$, our SCFT calculation shows that brushes in the presence of trivalent ions undergo microphase-separation and form lateral inhomogeneity. As shown in Figs. \ref{fig4}a and \ref{fig4}b, nearby brushes are pinned to micelles due to correlation-induced attraction, which further assemble into hexagonal pattern. We found that the cylindrical bundles proposed by previous scaling theory is unstable \cite{Brettmann_2017}. Our theoretical predictions are in good agreement with the observations from AFM images and simulations \cite{Yu_2017,Yu_2018,Jackson_2017,Tan_2023}.

The micro-phase separated morphologies of PE brushes are versatile. At high grafting densities (e.g. $\sigma=0.1$nm$^{-2}$), lateral inhomogeneities disappear; the segregation can be induced in the normal direction instead, as shown in Figs. \ref{fig4}c and \ref{fig4}d. Oscillatory structure appears inside the collapsed layer if ion correlations are enhanced by reducing the radius of the trivalent cation from $2.5\mathring{\rm A}$ to $1.5\mathring{\rm A}$. Similar ion-correlation induced oscillation has also been reported in the trivalent salt solutions near a charged surface \cite{Agrawal_2024}.

In this letter, we develop a new theory which systematically incorporates electrostatic fluctuations into the SCFT for polymers. Compared to existing work, our theory fully captures the coupling between ion correlations and polymer conformation without any preassumption on the chain structure and is particularly powerful in modeling inhomogeneous ion-containing polymers where ionic strength or dielectric permittivity is spatially varying. Applied to PE brushes, the theory predicts that ion correlations induce a non-monotonic change of the brush height: collapse followed by reexpansion. Our theoretical predictions are in quantitative agreement with the experimental data reported in the literature. We perform the scaling analysis to elucidate the essential role of the electrostatics as the competition between the repulsive osmotic pressure due to translational entropy and the attraction induced by ion correlations. Four regimes are identified in the height response: osmotic, salted, collapsed and neutral. We clarify the absence of causal relationship between the brush collapse-reexpansion and the charge inversion practically defined as the reversal of the surface electrostatic potential. Strong ion correlations also induce microphase separation in PE brushes. Depending on the parameters, the structural inhomogeneities can be either in the lateral direction as pinned micelles or in the normal direction as oscillatory layers. The easy access of our theory to the high dimensional calculation enables us to explore the rich morphological behaviors of PE brushes and other ion-containing polymer systems \cite{Sing:2013aa,Sing_2014,Gurovitch_1999,Dobrynin_2000,Winkler_2006,Mei:2006vx,Kegler:2008ue}.
Our theory can also be easily generalized to PEs with various chain architectures, copolymer compositions, charge patterns and dielectric inhomogeneity \cite{Yokokura_2023,Chen_2009,Pezeshkian_2012}.
Furthermore, it is important to note that non-monotonic behaviors triggered by multivalent ions have also been reported in the reversal of electro-osmotic flow and the attractive force between like-charged surfaces \cite{Heyden_2006,Besteman_2004,Agrawal_2022,Agrawal_2023,Agrawal_2024,Angelini_2003}, which reveals the universality of the salt concentration dependence of the electrostatic correlation.


Acknowledgment is made to the donors of the American Chemical Society Petroleum Research Fund for partial support of this research. This research used the computational resources provided by the Kenneth S. Pitzer Center for Theoretical Chemistry.



\bibliography{Tex_Refs}

\end{document}